\documentclass[prapplied,amssymb,amsmath,superscriptaddress,twocolumn]{revtex4-1}

\usepackage{graphicx,enumerate,mhchem,siunitx,gensymb,color}

\newcommand{\hl}[1]{\textcolor[RGB]{255,0,0}{#1}}
\renewcommand{\hl}[1]{#1} 

\graphicspath{{pix/}}
\usepackage[english]{babel}

\begin{document}

\title[]{Three-dimensional cooling of an atom beam source for high-contrast atom interferometry}

\author{J.~M. Kwolek}
\affiliation{NRC Postdoctoral Associate, U.S. Naval Research Laboratory, Washington, D.C. 20375}

\author{C.~T. Fancher}
\affiliation{The MITRE Corporation, McLean, Virginia 22102}

\author{M. Bashkansky}

\author{A.~T. Black}
\affiliation{Optical Sciences Division, U.S. Naval Research Laboratory, Washington, D.C. 20375}

\date{\today}

\begin{abstract}

We present a compact, two-stage atomic beam source that produces a continuous, narrow, collimated and high-flux beam of rubidium atoms with sub-Doppler temperatures in three dimensions, which features very low emission of near-resonance fluorescence along the atomic trajectory. The atom beam source originates in a pushed two-dimensional magneto-optical trap (2D$^+$ MOT) feeding a slightly off-axis three-dimensional moving optical molasses stage that continuously cools and redirects the atom beam. \hl{The capture velocity of the moving optical molasses is deliberately chosen to be low, $\sim 3$~m/s, to reduce fluorescence, and the cooling light is detuned by several atomic linewidths from resonance to reduce the absorption cross-section of cooling-induced fluorescence.} Near-resonance light from the 2D$^+$ MOT and the push beam does not propagate to the output atomic trajectory due to a $10\degree$ bend in the atomic trajectory.  The atomic beam emitted from the two-stage source has a flux up to $1.6(3)\times 10^9\;\textrm{atoms/s}$, with an optimized temperature of $15.0(2)$~\si{\micro K}. \hl{We employ continuous Raman-Ramsey interference measurements at the atom beam output to study the sources of decoherence in the presence of continuous cooling, and demonstrate that the atom beam source effectively preserves high fringe contrast even during cooling.} This cold-atom beam source is appropriate for use in atom interferometers and clocks, where continuous operation eliminates dead time, the slow atom beam velocity (6~-~16~m/s) improves sensitivity, the narrow 3D velocity distribution improves fringe contrast, and the low reabsorption of scattered light mitigates decoherence caused by the continuous cooling process. 

\end{abstract}

\maketitle

\section{Introduction}
Continuous-beam sources of laser-cooled atoms \cite{hansch_cooling_1975,lett_observation_1988,weiss_optical_1989,dalibard_laser_1989} are of interest in applications such as atom-based sensors and clocks because they increase measurement bandwidth and eliminate dead time compared with pulsed cold-atom sources. \hl{Compared with hot-atom beam sources, the slower mean velocity and narrower velocity distribution of cold atom beams both increase the interrogation time $T$ and increase the fringe contrast of atom interferometers and clocks, resulting in improved measurement sensitivity \cite{kasevich_measurement_1992,muller_compact_2009,luo_contrast_2016}.} Cold-atom sensors that operate periodically rather than continuously can undersample signals and noise, leading to the Dick effect in clocks \cite{dick_local_1987} and inertial navigation errors for accelerometers and gyroscopes \cite{jekeli_navigation_2005,cheinet_measurement_2008,joyet_theoretical_2012,dutta_continuous_2016,fang_metrology_2015}. 

Several types of continuous beam sources with cooling in one or more dimensions have been demonstrated, including slowed thermal beams \cite{phillips_laser_1982, park_generation_2002}, 2D magneto-optic traps (2D MOTs) \cite{riis_atom_1990,berthoud_continuous_1998,dieckmann_two-dimensional_1998}, continuous beams extracted from 3D MOTs \cite{lu_low-velocity_1996,kohel_generation_2003,feng_observation_2015}, continuous 3D moving optical molasses (3D OM) in an atomic vapor \cite{berthoud_bright_1999}, moving 3D OM fed by a 2D MOT \cite{castagna_novel_nodate,devenoges_improvement_2012}, \hl{MOT incorporating moving OM \cite{gorkhover_2008,basudev_2016}}, and continuous 2D degenerate Raman sideband cooling \cite{castagna_collimation_2004,di_domenico_laser_2004}. \hl{Additionally, alkaline earth metals like strontium can be cooled on multiple transitions, enabling the creation of cold atom beams from multiple stages of Doppler cooling \cite{bennetts_2017,chen_2019}.}

Continuous or interleaved cold-atom interferometer or clock measurements \cite{takase_precision_2008,devenoges_improvement_2012,biedermann_zero-dead-time_2013,meunier_stability_2014,dutta_continuous_2016,savoie_interleaved_2018} can increase measurement bandwidth and eliminate dead time, but require that cooling and trapping of atoms occur simultaneously with atomic interference. Therefore, near-resonance cooling light induces fluorescence in some atoms while nearby atoms are in a quantum superposition state. Photon scattering from an atom in a superposition state can induce atomic decoherence \cite{chapman_photon_1995,davis_raman_2016,black_decoherence_2020} that causes a loss of fringe contrast and decreases measurement sensitivity. Additionally, fluorescence emitted along the atomic trajectory can induce ac Stark shifts that perturb measurements \cite{davis_raman_2016,meunier_stability_2014,savoie_interleaved_2018}. Decoherence due to atomic fluorescence has been observed in a number of studies of atom interferometry \cite{dutta_continuous_2016,meunier_stability_2014,savoie_interleaved_2018, black_decoherence_2020}. This problem is most significant in interferometers with long $T$ and compact size because the aggregate decoherence increases with time spent in the superposition state and decreases with distance from the source of fluorescence.

Various methods have been employed to reduce the effect of scattered near-resonant cooling light on atomic interference. Decoherence can be mitigated by eliminating the optical path between the cooling region and the interferometer region - for example, by employing gravity-induced curvature of the atomic trajectory to separate the atomic trajectory from optical paths \cite{devenoges_improvement_2012,xue_continuous_2015} - however, this method is inapplicable for dynamic measurement, in free-fall, or in very compact instruments. Interleaving of periodic measurements in physically separate locations \cite{biedermann_zero-dead-time_2013,schioppo_ultrastable_2017} also eliminates decoherence at the cost of increased size and complexity while raising the potential for additional dynamics-induced errors. To date, there has been little systematic study of the influence of continuous laser cooling on decoherence in quantum coherent measurements such as atomic clocks and atom interferometers. In this work we perform such a study and demonstrate a method of mitigating this decoherence.

In addition to continuous operation and mitigation of fluorescence effects, a number of features are desirable for the employment of a continuous-beam cold-atom source in atom interferometers or clocks on moving platforms: (1) high-flux atom output to obtain a large signal-to-noise ratio for either technical-noise-limited or quantum-projection-noise-limited \cite{itano_quantum_1993} operation; (2) narrow velocity distribution in three dimensions to increase interference fringe contrast and reduce errors induced by dynamics \cite{lan_influence_2012,stockton_absolute_2011,zimmermann_representation-free_2019}; (3) precise knowledge and control of the mean atomic velocity to determine $T$; (4) spatially narrow transverse atom beam size to reduce position-related errors; (5) low background of uncooled atomic vapor that can induce collisional decoherence \cite{uys_matter-wave_2005} and background fluorescence; \hl{(6) ability to operate in arbitrary spatial orientation and under dynamics. Each of the examples of continuous atom sources given above focuses on one or more of these goals, and the atom source presented in this work meets all of them.}

In this work, we demonstrate and characterize a compact, 3D-cooled atom beam source that is designed to enable continuous atom interferometer or clock operation to eliminate dead time and provide high fringe contrast, while also minimizing the decoherence that scattered cooling light can produce. The source combines several features: continuous, high-flux ($1.6(3) \times 10^9$~atoms/s) atom output; continuous sub-Doppler cooling ($15-30\;\mu$K) of atoms in three dimensions; precisely controlled mean atom velocity (6 to 16 m/s); narrow transverse atom beam size \hl{(Gaussian $e^{-1/2}$ radius $1.35(5)\;\textrm{mm}$)}; low background of uncooled atomic vapor; and very low emission of light at or near atomic resonance frequencies along the atomic trajectory. As a demonstration of the low decoherence due to cooling-induced fluorescence, we perform continuous Ramsey fringe measurements at a distance of 7.5 cm from the final atom beam cooling stage. \hl{We characterize experimentally the decoherence of Ramsey interference due to continuous cooling and compare it with other sources of decoherence.}

\section{Experimental Design and Model}\label{exdesign}

The compact cold-atom beam source consists of a longitudinally pushed \ce{^87Rb} 2D MOT (2D$^+$ MOT) \cite{dieckmann_two-dimensional_1998} that feeds a narrow beam of 2D-cooled atoms into a continuously operating moving 3D OM. The 2D$^+$ MOT provides a high-flux beam of atoms that is narrow in transverse size and is cooled close to the Doppler limit in transverse velocity. The atom beam propagates through a pinhole into the 3D OM region. The 3D OM stage further cools the atoms to sub-Doppler temperatures by polarization gradient cooling in three dimensions, and it continuously redirects the mean atomic velocity by an angle $\theta=10\degree$ compared with the output axis of the 2D$^+$ MOT. 

The use of the continuous moving 3D OM following the 2D$^+$ MOT stage reduces the degradation in downstream quantum coherent processes for a number of reasons. \hl{The redirection of the atom beam makes it possible to spatially separate the atom trajectory from both the push beam and 2D$^+$ MOT fluorescence passing through the pinhole. Differential vacuum pumping due to the pinhole between the two regions ensures that there is very little warm atomic vapor in the 3D OM stage, resulting in reduced background collisions and fluorescence.} Because of the small angle $\theta$ between the output axes of the 2D$^+$ MOT and 3D OM, the capture velocity needed to 3D-cool and redirect the 2D$^+$ MOT beam, approximately 3~m/s, is an order of magnitude smaller than that needed to produce a high-flux 2D$^+$ MOT output, and is also much smaller than would be needed to stop the 2D$^+$ MOT atom beam in a stationary 3D MOT or OM. The photon scattering rate required to achieve this capture velocity, and thus the rescattered optical power, is therefore reduced. Finally, the 3D OM region operates at a larger detuning from atomic resonance than the 2D$^+$ MOT, $\Delta_{OM} = 6 \Delta_{2D} =-6 \Gamma_0$, causing rescattered cooling light to have a lower cross-section for photon reabsorption by a factor of $(\Gamma_0^2+4 \Delta_{OM}^2)/(\Gamma_0^2+4 \Delta_{2D}^2) \approx 29$, where $\Gamma_0$ is the natural linewidth of the atomic cooling transition, $\Delta_{OM}$ is the central detuning of the 3D OM, and $\Delta_{2D}$ is the detuning of the 2D$+$ MOT.

\begin{figure}
	\includegraphics[width=\linewidth]{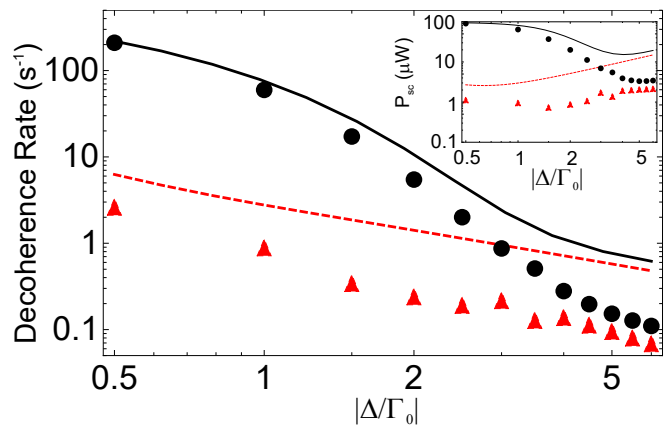}
	\caption{\label{fig:decoherenceplot} Numerical models of decoherence rate caused by cooling-induced fluorescence as a function of detuning from atomic resonance, for \ce{^87Rb} atoms located 7.5 cm from the end of a 2D transverse laser cooling region. The transverse capture velocity is held constant at either 40 m/s (black) or 4 m/s (red) by adjusting cooling laser power as the detuning is varied. The cooling length is 5 cm, longitudinal velocity is 10 m/s, and the atom flux is $2 \times 10^9$~atoms/s. The solid black (dashed red) line shows the result for a 2-level Doppler cooling model with 40~m/s (4~m/s) transverse capture velocity. The black circles (red triangles) show the result for a numerical model of $\sigma^+/\sigma^-$ polarization gradient cooling with 40~m/s (4~m/s) transverse capture velocity. Inset: Fluorescence power as a function of detuning for the same set of models and parameters.
	}
\end{figure}

A simple 2-level Doppler cooling and fluorescence model illustrates the effectiveness of this approach: consider the transverse cooling of a narrow atom beam with flux $\Phi$ at uniform longitudinal velocity $v_x$ in a cooling length $L$, with steady state atom number $N=\Phi \frac{L}{v_x}$. The scattered power due to the cooling beams is $P_{sc}=N \Gamma_{sc} \hbar \omega$, where  $\Gamma_{sc}$ is the scattering rate due to the cooling beams, $\hbar$ is the reduced Planck constant and $\omega$ is the angular frequency of the cooling light. The resulting spontaneous emission rate for downstream atoms along the output axis at a distance $r$ from the atom beam output due to isotropically rescattered cooling light, is
\begin{equation}
\label{eqn:spontaneousemission}
\Gamma_{sp} = \frac{P_{sc} }{4 \pi I_{sat} (L r + r^2)}\frac{\frac{\Gamma_0}{2}}{1+(\frac{2 \Delta}{\Gamma_0})^2}
\end{equation} where $\Delta$ is the detuning of the rescattered light from atomic resonance and $I_{sat}$ is the saturation intensity. In the frame of reference of the atom beam and in the limit of low saturation \cite{mollow_power_1969}, $\Delta$ is equal to the detuning of the incident cooling laser beams up to the Doppler shift due to the transverse atomic motion and the recoil shift.
We find that the scattered optical power $P_{sc}$ varies relatively weakly as a function of detuning for constant capture velocity (see inset to Fig.~\ref{fig:decoherenceplot}). For the purpose of the model we define the capture velocity as the maximum transverse velocity at the input to the cooling region that results in a final velocity below 1~cm/s at the output of the cooling region, as obtained from numerical solutions to the equations of motion for 1D Doppler cooling\cite{metcalf_laser_2003}. Increasing the detuning $\Delta$ while keeping $P_{sc}$ approximately unchanged therefore reduces the rate of decoherence in downstream atoms due to scattered cooling light, as shown in Fig.~\ref{fig:decoherenceplot}. 

\begin{figure*}
	\includegraphics[width=\linewidth]{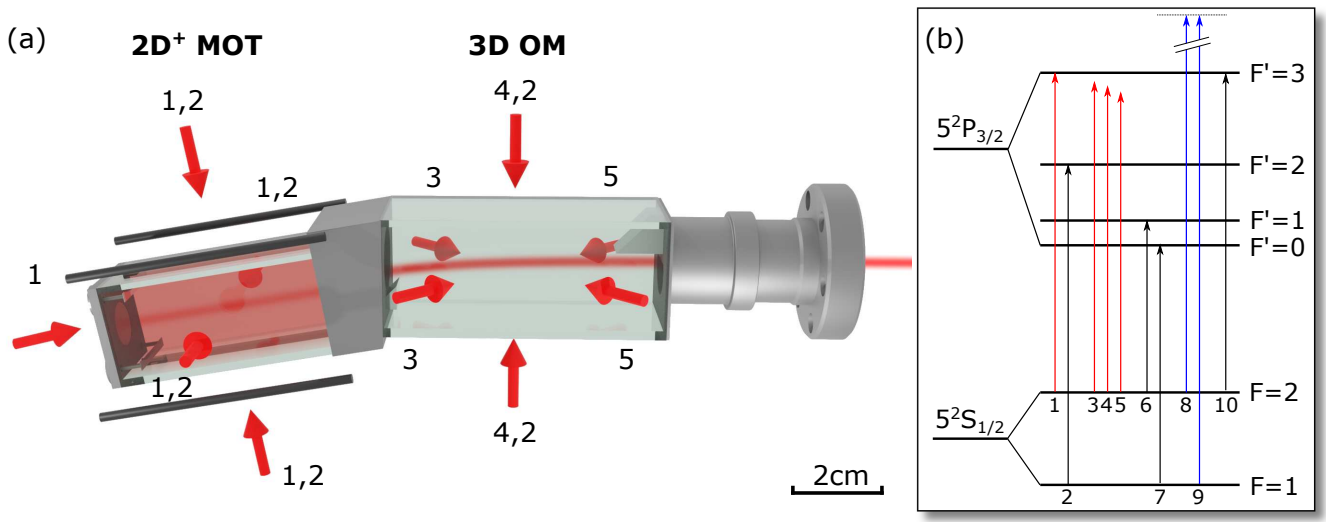}
	\caption{\label{fig:apparatus} The atom beam source vacuum cell (a) consists of two cells anodically bonded to a silicon spacer that imposes a $10\degree$ angular offset between the cells. The cell shown on the left is surrounded by four permanent magnets (magnetic shielding not shown), used in conjunction with lasers to create a 2D$^+$ MOT which feeds atoms through a pinhole in the silicon spacer. The cold atoms travel into the 3D OM region, where they are redirected and cooled further. The atom trajectory is depicted with a diffuse red line inside the cell. Lasers are depicted as red arrows. The labels above each grouping of lasers indicate the frequencies in each laser beam, and correspond with a label on the level diagram (b) which is not drawn to scale. In the apparatus depiction (a), the cooling (1), repump (2), and forward/vertical/backward OM (3/4/5 respectively) laser frequencies are shown. Two mirrors in the second cell reflect light emitted from the 2D MOT chamber. Additional lasers not shown in (a) are depicted in (b). We optically pump into $F=1$, $m_F=0$ ground state with a depumping laser (6) and a linearly polarized optical pumping beam (7). We coherently manipulate the atomic state by driving stimulated Raman transitions (8,9), and read out the final state population in the $F=2$ ground state with an on-resonance laser (10).
	}
\end{figure*}

The inclusion of polarization gradient cooling forces \cite{dalibard_laser_1989,nienhuis_operator_1991} in the analysis markedly improves the predicted effectiveness of the cooling at large detuning in comparison with the simple 2-level model above. We follow the semiclassical polarization gradient force analysis, which is described in detail in Ref.~\cite{nienhuis_operator_1991}. \hl{Atomic motion is treated classically while the evolution of the atomic internal-state density matrix $\sigma (t)$ is described by the optical Bloch equations (OBEs) for the \ce{^87Rb} $F=2 \to F'=3$ transition. We subdivide $\sigma (t)$ into submatrices $\sigma_{ee}(t)$ and $\sigma_{gg}(t)$, describing the excited and ground states respectively, and optical coherences $\sigma_{eg}(t)$ and $\sigma_{ge}(t)$. The numerical solution to the OBEs provides the time evolution of the density matrix for an atom moving at constant velocity $v_z$ through a counterpropagating 1D pair of laser beams along the $z$ axis. We evolve the system until a steady-state oscillatory solution is obtained for $\sigma (t)$. The positive-frequency part of the atomic dipole moment is determined by $\bf{\mu}_+(t) = \mathrm{Tr} (\sigma_{eg}(t) \mu_{ge})$, where $\mu_{ge}$ is the atomic electric dipole operator. The instantaneous force is then }
\begin{equation}
\label{eqn:PGCforce}
\hl{f_z(t)=2\mathrm{Re}\left(\bf{\mu}_+(t)\cdot\frac{d}{dz} \bf{E}_-\right)}
\end{equation}
\hl{where $\bf{E}_-$ is the negative frequency component of the electric field vector. The average force over an optical wavelength is used to determine curves of acceleration versus velocity, which we integrate numerically to obtain velocity versus time and thereby determine the optical molasses capture velocity. The simulation also evaluates the cooling-induced fluorescence rate as}
\begin{equation}
\label{eqn:PGCfluorescence}
\hl{\Gamma_{sc} = \Gamma_0 \mathrm{Tr}(\sigma_{ee})}\,.
\end{equation} The resulting decoherence rate from Eq.~\ref{eqn:spontaneousemission} at constant capture velocity in Fig.~\ref{fig:decoherenceplot} shows a reduction by nearly an order of magnitude at large detunings compared with the 2-level model.

\section{Apparatus}

A schematic diagram of the cold atom beam source is shown in Fig.~\ref{fig:apparatus}a. The 2D$^+$ MOT is produced by two perpendicular pairs of counterpropagating circularly polarized beams with $\Delta_{2D}=-\Gamma_0$ relative to the $F=2\to F'=3$ transition in the $S_{1/2}\to P_{3/2}$ $\textrm{D}_2$ transition line of \ce{^87Rb}. Repump light on the $F=1\to F'=2$ transition pumps atoms out of the $F=1$ ground-state. A longitudinal push beam at the 2D$^+$-MOT cooling frequency enhances the atomic flux by an order of magnitude. The \ce{^87Rb} atomic level structure and laser frequencies are shown in Fig.~\ref{fig:apparatus}b, with the cooling and repump light labeled as 1 and 2 respectively. The 2D$^+$ MOT is loaded from a hot background vapor provided by a Rb dispenser. A two-dimensional quadrupole magnetic field is created by four permanent magnets located near the corners of the 2D$^+$ MOT vacuum cell. To reduce magnetic perturbations in the other regions of the vacuum cell, an annealed mumetal shield is situated around the 2D$^+$ MOT region. 

The 3D OM is created by three pairs of counter-propagating laser beams of orthogonal polarization. The geometry and frequencies of the 3D OM beams are chosen to cool the atoms into a frame of reference moving at a 10$\degree$ angle from the 2D$^+$ MOT output axis, in the (1,1,0) direction with respect to the 3D OM cooling laser propagation directions. Polarization gradient cooling (PGC) can operate with either lin$\perp$lin or $\sigma^+$/$\sigma^-$ light in each counterpropagating pair of beams. We find no substantial difference in the atomic flux and temperature between the two cooling schemes, possibly because of magnetic field inhomogeneities discussed later. We use lin$\perp$lin light in the 3D OM throughout the experiments discussed here. The mean detuning of the laser beams in the 3D OM is $\Delta_{OM}=-6\Gamma_0$. The vertical OM beam (4 in Fig.~\ref{fig:apparatus}b) is retroreflected, while the other two OM beam pairs (3 and 5 in Fig.~\ref{fig:apparatus}b) are symmetrically frequency shifted to cool into the moving frame of reference. This frequency difference is tuned to control the output velocity of the 3D OM. A repump beam is overlapped with the vertical OM beam.

The compact vacuum cell for the two-stage beam source was manufactured by ColdQuanta, Inc. Its length is 12 cm plus an additional 4 cm for the transition from the glass cell to a steel CF flange used for testing purposes. The 2D$^+$ MOT chamber is a glass cell anodically bonded to a silicon spacer, which incorporates a $10\degree$ wedge angle and a 0.75 mm diameter pinhole. The 3D OM chamber is a second glass cell anodically bonded to the same spacer. A pair of mirrors at the beginning and end of the 3D OM region reflects light transmitted from the 2D$^+$ MOT region out of the cell, including the push-beam light and scatter from the thermal background atoms. For testing purposes, the atom beam source is attached to a Kimball Physics vacuum chamber not shown in Fig.~\ref{fig:apparatus}a.

\hl{To optimize PGC we null the magnetic field along each axis independently in the 3D OM region using three orthogonal pairs of Helmholtz coils situated around the 3D OM cell. Final optimization of the magnetic field is obtained by minimizing the temperature of atoms leaving the 3D OM region.} The final temperature is unchanged over a $\sim 20\;\textrm{mG}$ range of magnetic fields along each axis.

We characterize the atomic beam velocity distribution and decoherence with Doppler-sensitive and Doppler-insensitive stimulated Raman processes respectively. The stimulated Raman transition is driven using a single-photon detuning $1.1\;\textrm{GHz}$ above the \ce{^87Rb} cycling transition to avoid single-photon resonant excitation. The laser frequencies used to drive Raman transitions are labeled as 8 and 9 in Fig.~\ref{fig:apparatus}b. The first laser frequency (8) is stabilized by saturated absorption spectroscopy to a vapor cell on the \ce{^85Rb} ($F=3\to F'=4$) $\textrm{D}_2$ transition. The second laser frequency (9) is created by phase modulation of (8) using an electro-optic phase modulator (EOM) at the ground-state separation of \ce{^87Rb} \hbox{($\sim 6.834\;\textrm{GHz}$)}. A spectrum of stimulated Raman transitions between different magnetic sublevels is observed by scanning the EOM modulation frequency across the two-photon resonances. \hl{To eliminate frequency shifts due to the differential ac Stark shift, we adjust the Raman beam modulation depth so that the measured clock transition ($F=1,m_F=0\to F=2,m_F=0$) frequency is stationary with respect to Raman beam optical power.} A typical Doppler-free Raman peak width is $40\;\textrm{kHz}$ FWHM, consistent with the Fourier transform limit by the transit time of atoms across the Raman beam.

\begin{figure}
	\includegraphics[width=\linewidth]{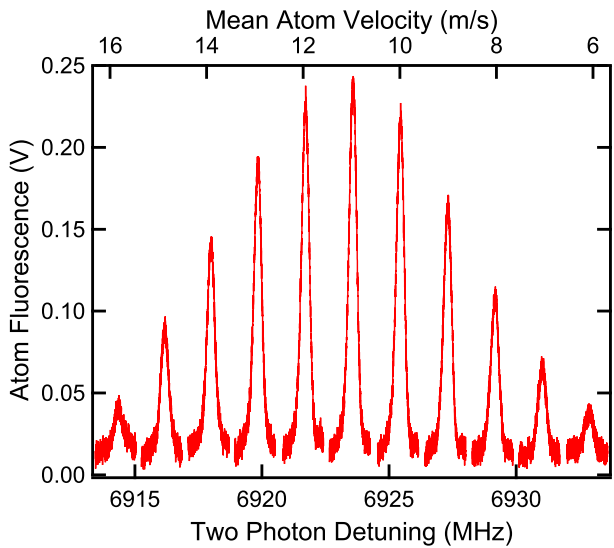}
	\caption{\label{fig:dsensitive} \hl{Doppler-sensitive Raman spectra are used to measure the velocity at 45 degrees to the atom beam propagation direction, providing an estimate of both mean atom beam velocity and 1D temperature along the Raman beam propagation axis.} Each peak here represents a different moving OM frequency shift in the secondary cooling stage. The flux of cold atoms is optimized for each OM detuning by changing the push-beam intensity. The peak height is determined by the atom velocity as well as the atomic flux, and the atom beam flux is measured independently. 
	} 
	
\end{figure}

To excite Doppler-sensitive Raman transitions, counterpropagating laser beams with orthogonal linear polarizations are used. The two beams are derived from a common laser source and individually phase-modulated at a frequency away from the Doppler-free Raman resonance. One of the two beams is frequency-shifted using an acousto-optic modulator (AOM). We individually zero the differential ac Stark shift of each beam by the procedure described above. The Doppler-sensitive Raman transitions are driven using the carrier from one beam in the counterpropagating pair and the first-order blue sideband of the other. The phase modulation frequency is chosen to meet the Doppler-sensitive Raman resonance condition on the clock transition, accounting for the AOM frequency shift and the Doppler shift due to the projection of the atom beam velocity onto the direction of propagation of the Raman beam pair. For Raman beams at $45\degree$ to the atom beam direction, this modulation frequency ranges from $6.91\;\textrm{GHz}$ to $6.95\;\textrm{GHz}$, depending on the chosen atom beam velocity. Typical Doppler-sensitive Raman transition spectra are shown in Fig.~\ref{fig:dsensitive} for many different atom beam velocities.

When performing Raman spectroscopy on the clock transition, we prepare the atoms in the $F=1, m_F=0$ ground state. Since the 3D OM region is illuminated with repump light, the atoms enter the final chamber primarily in the F=2 ground state and primarily in the higher-lying magnetic sublevels. To optically pump the atom beam to $F=1, m_F=0$, we apply an on-resonance laser on the $F=2\to F'=1$ transition (6 in Fig.~\ref{fig:apparatus}b), and linearly polarized optical pumping light on the $F=1\to F'=0$ transition (7 in Fig.~\ref{fig:apparatus}b). Measurements of Doppler-free Raman spectra indicate that 94\% of atomic population is in the $F=1, m_F=0$ state following optical pumping. Optical pumping increases the signal size in the measurement of Raman-Ramsey fringes on the clock transition, described below, by a factor of $\sim 8$.

To perform state-dependent readout of the atomic beam, we illuminate the atoms using a laser on resonance with the $\textrm{D}_2$ ~line cycling transition (10 in Fig.~\ref{fig:apparatus}b). The atomic fluorescence in the detection beam is collected by an avalanche photodiode with a two-achromat light-collection system. With this measurement scheme, we quantify the atomic flux in the F=2 ground state.

\section{Characterization}

The two-stage atom beam source is designed primarily for use as a continuous source of 3D-cooled atoms for atom interferometry. In this section, we characterize the atom beam in the context of figures of merit that are important in atom interferometry. We divide the section into two parts. First, we characterize the atomic beam, including measurements of temperature and flux. In the second section, we characterize the many sources of decoherence, including decoherence induced by the continuous operation of the cold-atom beam source.

\subsection{Atomic beam}

We measure the 3D velocity distribution of the atom beam via the spectra of Doppler-sensitive Raman transitions at a variety of angles to the atom beam \hl{\cite{kasevich_atomic_1991}. In our continuous beam measurement, the time-dependence of the Raman transition 2-photon Rabi frequency is $\Omega(t) = \Omega_0 \exp (-2 v_x^2 t^2/w^2)$ where $\Omega_0$ is the peak Rabi frequency, $w$ is the Gaussian Raman beam waist and $v_x$ is the atomic velocity along the direction perpendicular to Raman beam propagation. In a 2-level approximation without damping, with atomic population initialized in the $F=1, m_F=0$ state, the single-atom population of the $F=2, m_F=0$ state following a Doppler-sensitive Raman $\pi$ pulse is 
\begin{equation}
\label{eqn:transitionprobability}
\left|c_2 (\delta) \right|^2=\exp\left[-\left(\frac{(\delta+2 \vec{k} \cdot \vec{v}) w}{2 v_x}\right)^2\right]
\end{equation}
where $\delta$ is the 2-photon detuning of the driving fields from atomic resonance, $\vec{k}$ is the wavevector of the Raman beam, $\vec{v}$ is the atomic velocity, and $\Omega$ is the 2-photon Rabi frequency. Integration of the single-atom population over the 1D Maxwell-Boltzmann atomic velocity distribution along the $\vec{k}$ direction results in a total population 
\begin{equation}
\label{eqn:averagetransitionprobability}
\rho_2 (\delta) \propto \exp \left[ -\frac{1}{2} \left(\frac{\delta+ 2 k v_0}{2 k \sigma_v}\right)^2 \right]
\end{equation}
in the limit $v_x/w << 2 k \sigma_v$, where $v_0$ is the mean velocity along $\vec{k}$ and $\sigma_v$ is the 1D standard deviation of velocity along $\vec{k}$. Measurement of $\rho_2$ as a function of $\delta$ thereby provides a measurement of the velocity distribution along the direction of propagation of the Raman beam.} 

\begin{figure}
	\includegraphics[width=\linewidth]{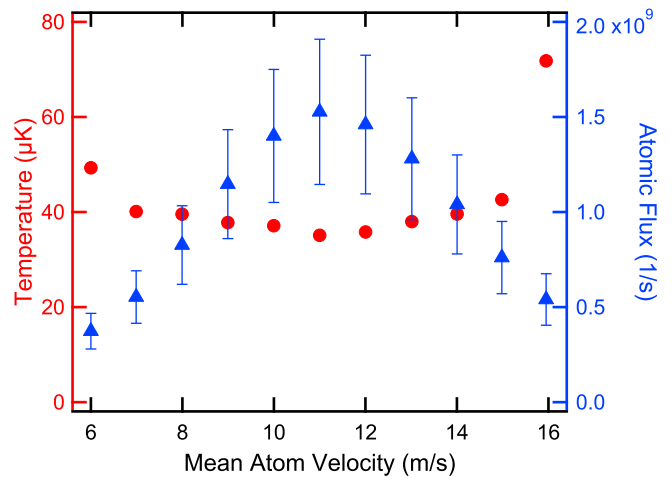}
	\caption{\label{fig:tempvsvel} The atomic flux (blue triangles) and atomic temperature (red circles) vary as a function of the velocity of the moving OM frame of reference. The atomic temperature is measured on a $45\degree$ angle from the direction of atom travel. The apparatus produces a flux of $>5\times 10^8\;\textrm{atoms/s}$ with temperatures $<40\;\mu\textrm{K}$ between $7\;\textrm{m}/\textrm{s}$ and  $15\;\textrm{m}/\textrm{s}$, with a peak flux of $1.5\times 10^9\;\textrm{atoms/s}$ which corresponds to a minimum temperature of $35(3)\;\mu\textrm{K}$. The temperature error bars are smaller than the points.
	}
	
\end{figure}

In our experiment, the temperature measurement is limited by a lower bound of 500 nK given by the Fourier transform of the Raman pulse profile, which is well below our lowest achieved atom temperatures. \hl{We orient the Raman beams at a $45\degree$ angle from the longitudinal direction of travel, and thereby measure the temperature in a linear combination of longitudinal and transverse dimensions. To supplement this measurement, we also perform sequential measurements of the temperature in each transverse dimension independently by reorienting the Raman beam direction.} The typical atomic temperature is determined to be $35(3)\;\mu\textrm{K}$ in the $45\degree$ off-axis longitudinal-horizontal dimension, and $30(2)\;\mu\textrm{K}$ and $28(2)\;\mu\textrm{K}$ in the vertical and horizontal transverse dimensions respectively. Each of these temperatures is measured with the atoms at a mean velocity of $11\;\textrm{m}/\textrm{s}$. Fig.~\ref{fig:tempvsvel} shows a measurement of temperature as a function of mean atom beam velocity, which gives slightly higher minimum temperatures due to imperfect optimization of magnetic fields. As we discuss below, further temperature optimization subsequent to these measurements has led to lower measured temperatures.

We corroborate the measurement of longitudinal atom temperature and mean velocity by measuring the atomic time-of-flight signal. A plug-beam situated at the end of the 3D OM region pushes the atoms away from a longitudinally-narrow on-resonance detection beam, located $30\;\textrm{mm}$ further along the atom path. We modulate the plug-beam off for a short period of time and let a small pulse of atoms through, which continues to the detection beam. We measure the temporal profile of the resulting fluorescence. The minimum temporal width of the fluorescence pulse is limited by the mean atom velocity, the atomic pulse size, and the detection beam size. The resulting time resolution of the measurement places a lower bound on detectable temperature of $50\;\mu\textrm{K}$ for longitudinal velocities $v_x>10\;\textrm{m}/\textrm{s}$. The temperature values measured in this way are roughly consistent with those measured by Doppler-sensitive Raman spectroscopy within the limited temporal resolution of the time-of-flight technique. For example, time-of-flight measurements indicate a longitudinal temperature of $44(3)\;\mu\textrm{K}$ at a mean velocity of $8\;\textrm{m}/\textrm{s}$, similar to Doppler-sensitive Raman measurements at the same velocity taken under similar conditions (See Fig.~\ref{fig:tempvsvel}).

While the majority of the characterization of the atom beam source in the present work was conducted with minimum measured 3D atom beam temperatures of $\sim 30$~\si{\micro K}, subsequent optimization of OM beam pointing and magnetic field homogeneity has resulted in regularly achievable temperatures of 15.0(2)~\si{\micro K} for atom velocity 11~m/s. As we discuss below, further optimization is expected to achieve still lower 3D atom beam temperatures. 

The absolute atomic flux is determined by illuminating the atom beam with a perpendicular, continuous, on-resonance probe beam and measuring the fluorescence with a calibrated camera. The probe light is modulated to include a repumping sideband to ensure an accurate measurement of the atom number. We select a cross-section of the atomic beam, and with the calculation of solid angle, calibration, and mean atom velocity we determine the peak atomic flux to be $1.6(3)\times 10^9\;\text{atoms}/\text{s}$ at $v_x = 11\;\textrm{m}/\textrm{s}$. Without changing the mean 3D OM detuning, the flux changes as a function of the 3D OM velocity as seen in Fig.~\ref{fig:tempvsvel}. This variation is expected due to the varying overlap between the velocity capture range of the 3D OM and the velocity distribution of the 2D$^+$ MOT atom beam as the 3D OM velocity is varied.

\subsection{Coherence and Scattered light}

 \begin{figure}
\includegraphics[width=\linewidth]{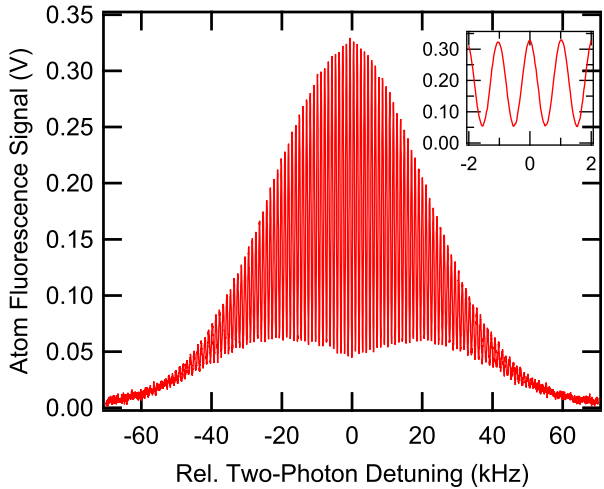}
	\caption{\label{fig:ramsey} Continuous Raman-Ramsey interference is observed using two laser beams driving Doppler-insensitive Raman $\pi/2$-pulses $11.0\;\textrm{mm}$ apart with atomic velocity 11~m/s. \hl{The fringes are produced by sweeping the common microwave frequency driving the two EOMs that produce phase modulation sidebands of the two Raman beams.} Inset: detail of the central fringe.
}

\end{figure}

To investigate the influence of steady-state cooling light on quantum coherence in the atom beam, we perform continuous Raman-Ramsey interferometry \cite{ramsey_molecular_1950,thomas_observation_1982,feng_observation_2015} on the cold atom beam at a location 7.5 cm from the 3D OM cooling laser beams. The Raman-Ramsey interference is induced by two spatially separated laser beams driving Doppler-insensitive $\pi/2$ pulses on the $F=1, m_F=0 \leftrightarrow F=2, m_F=0$ clock transition. The final measured $F=2$ state population $\rho_2$ is sensitive to the phase accumulated between the pulses. \hl{The fringes are produced by sweeping the common microwave frequency driving the two EOMs that produce phase modulation of the Raman beams driving the two $\pi/2$-pulses.} Initializing the atoms in the $F=1$, $m_F=0$ ground state and measuring the output population in $F=2$ state, the output population at the central fringe is 
\begin{equation}
\rho_2(\delta) = \bar{\rho_2}\left(1+C \cos\left(\frac{\delta \ell}{v_x}\right)\right),
\end{equation}
 where $\delta$ is the frequency difference between the laser phase modulation and the atomic resonance frequency, $\ell$ is the free evolution distance, $\bar{\rho_2}$ is the mean population of the central fringe and $C$ is the central fringe contrast. Fig.~\ref{fig:ramsey} shows an example of continuously measured Raman-Ramsey fringes taken at the atom beam output.

The Raman-Ramsey fringe contrast $C$ provides an estimate of the decoherence that occurs during the free evolution time. Sources of fringe contrast loss include spontaneous emission, inhomogeneity of Raman beam intensities, background gas collisions, and magnetic field inhomogeneity. Fringe contrast loss mechanisms related to driving field inhomogeneity and spontaneous emission induced by the Raman beams are largely insensitive to the pulse separation, while decoherence induced by scattered light, magnetic field inhomogeneity, and collisions depend on pulse separation.

\begin{figure}
	\includegraphics[width=\linewidth]{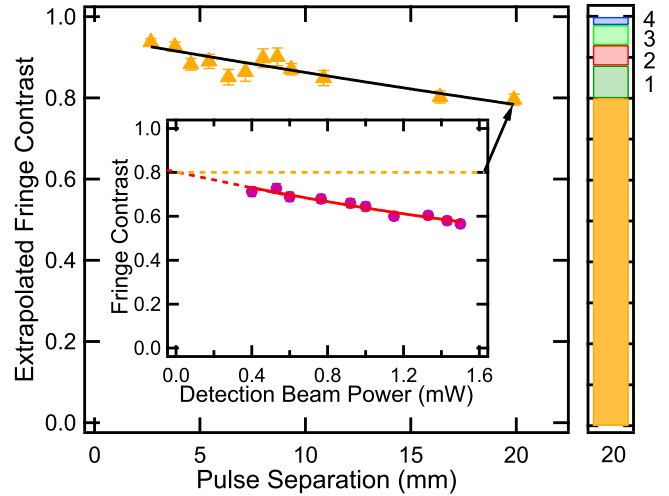}
	\caption{\label{fig:extrapcontrast} Raman-Ramsey fringe contrast as a function of free evolution length. A single-exponential fit to the length-dependent decay gives a contrast decay length constant of $110(3)\;\textrm{mm}$. \hl{The color bar (right) summarizes the measured sources of contrast loss at 20 mm pulse separation: (1, dark green) surface scattering and reflections: $8\%$; (2, red) length-independent fringe contrast loss due to Raman beam spontaneous emission and pulse area error: $4.9(1)\%$; (3, bright green) uncharacterized length-dependent fringe contrast loss, likely due to magnetic field inhomogeneities over the Ramsey region: $5\%$; (4, blue) rescatter from atoms in the OM: $2(2)\%$. The bottom portion of the bar indicates the remaining fringe contrast of 80\%.} Inset: The fringe contrast also degrades as a function of detection beam intensity. To provide the fringe contrast measurements in this figure, we fit the exponential decay of the fringe contrast as a function of detection beam power in order to infer the fringe contrast at 0 detection beam power.
}

\end{figure}

We measure fringe contrast as a function of $\ell$ to distinguish various decoherence mechanisms, as shown in Fig.~\ref{fig:extrapcontrast}. Some decoherence occurs due to scattering from the on-resonance detection beam, which is located $< 1$~cm away from the output of the Ramsey region. For the purposes of quantifying decoherence due to the cold-atom source alone, the fits shown are extrapolated to zero detection beam power (Fig.~\ref{fig:extrapcontrast} inset). \hl{By using a slightly longer test chamber allowing the probe beam to be placed at a greater distance from the Ramsey region, it will be possible to almost entirely eliminate decoherence due to the state-detection laser.}

Length-independent fringe contrast loss is measured by fitting the fringe contrast as a function of pulse separation (Fig.~\ref{fig:extrapcontrast}). The y-intercept of the exponential fit gives the length-independent contrast loss as $4.9(1)\%$. We calculate that half of this length-independent contrast loss is attributable to spontaneous emission induced by the Raman beams, and the remainder is probably due to imperfect Raman pulse area.

Here we summarize the contributions to decoherence at $\ell=20$~mm \hl{(summarized in the color bar in Fig.~\ref{fig:extrapcontrast})}. Once probe-induced decoherence is extrapolated to zero, the most significant observed source of fringe contrast degradation is surface scattering and reflections from the vacuum cell walls. At $\ell=20$~mm, a $10\%$ improvement in the fringe contrast occurs when the 2D MOT and 3D OM cooling beams are shuttered during the time spent by the atoms in the interference region. Some of this fringe contrast degradation is due to atomic fluorescence in the 3D OM, while the remainder is due to other scattering and reflection paths. To isolate atomic fluorescence as a source of decoherence, we measure the fringe contrast as a function of atom beam flux by varying the rubidium dispenser current. A linear fit of fringe contrast vs flux determines a $2(2)\%$ loss per $10^9$ atoms/s at $\ell=20$~mm. These measurements also indicate $5\%$ of additional length-dependent degradation unrelated to cooling light, which is probably due to magnetic field inhomogeneity across the Ramsey interference region.

As discussed in Sec.~\ref{exdesign}, the decoherence due to atomic fluorescence in the 3D OM decreases as a function of detuning. To confirm this, we measure decoherence at $\ell=20$~mm and compare operation of the 3D OM at $\Delta_{OM}=-\Gamma_0$ versus $\Delta_{OM}=-6 \Gamma_0$. To equalize capture velocity for the two detunings, we adjust the intensity in all three OM beam pairs by a factor of 150 between the two measurements. For  $\Delta_{OM}=-\Gamma_0$, we measure a loss of $53(8)\%$ fringe contrast per $10^9$ atoms/s due to atomic fluorescence, in contrast with $2(2)\%$ per $10^9$ atoms/s at $\Delta_{OM}=-6\Gamma_0$. This is consistent with the nominal predicted relationship between OM detuning and decoherence due to cooling-induced fluorescence for a given capture velocity as indicated in Fig.~\ref{fig:decoherenceplot}.

To compare the decoherence due to cooling-induced fluorescence in the two-stage atom beam source with the decoherence that would be expected due to a single-stage cold-atom source that cools directly from warm atomic vapor at high capture velocity, we measure the fluorescence power emitted on-axis from the 2D$^+$ MOT via the on-axis vacuum cell window and calculate the expected photon scattering rates based on optical power and detuning. Using a ray tracing model, we infer the fluorescence power emitted from the 0.75 mm diameter aperture at the 2D$^+$ MOT output. By modifying the detuning of the 2D$^+$ MOT cooling beams and the rubidium dispenser current, we differentiate between power scattered from the cold-atom beam, the hot atomic vapor, and surfaces in the 2D MOT cell.

At a distance of 7.5 cm from the 2D$^+$ MOT output, the inferred on-axis intensity due to fluorescence by the 2D$^+$ MOT cold atom beam alone is 250~nW/\si{cm^2}. The resulting calculated scattering rate at $\Delta_{2D} = -\Gamma_0$ for atoms in the $F=2$, $m_F=0$ state, assuming isotropic incident polarization, is 270~\si{s^{-1}}. By comparison, the measured decoherence rate due to cooling-induced fluorescence from the 3D OM based on Ramsey fringe contrast (Fig.~\ref{fig:extrapcontrast}) at a similar distance from the atom beam output is $10 \pm 10$~\si{s^{-1}}. The semiclassical polarization gradient cooling model \cite{nienhuis_operator_1991} is roughly consistent with this result within measurement error, predicting an on-axis fluorescence intensity from the 3D OM on the order of 10~nW/\si{cm^2} at a distance of 7.5 cm from the beam output and implying a decoherence rate of 0.5~\si{s^{-1}}. The two-stage cooling method thus provides an inferred improvement in fluorescence-induced decoherence rate by a factor of $\sim 500$ compared with the single-stage 2D MOT fluorescence. As illustrated in Fig.~\ref{fig:decoherenceplot}, we attribute this improvement to a combination of reduced capture velocity and increased detuning in the second cooling stage compared with typical single-stage beam source parameters.

\section{Discussion and Outlook}
To estimate the limits of potential performance of continuous inertial sensor and clock architectures based on the demonstrated cold-atom beam, we compute short-term noise at the quantum projection noise (QPN) limit \cite{itano_quantum_1993}. At the QPN limit, the phase noise of an atom interferometer per root-bandwidth is given by $\Delta \phi \approx (C \sqrt{\Phi})^{-1}$, irrespective of its operation in a pulsed or continuous mode. The atom source demonstrated here has a peak flux $\Phi \approx 1.5\times 10^9$~atoms/s, similar to typical 3D MOT loading rates, which implies similar performance at the QPN limit to pulsed cold-atom interferometers at equivalent $T$ and $C$. For ideal fringe contrast $C=1$, the resulting QPN-limited phase noise is $\Delta\phi \approx 26$~\si{\micro rad}/\si{\sqrt{Hz}}. Assuming two counterpropagating atom beams in a three-Raman-pulse Mach-Zehnder atom interferometer configuration with interrogation time $T=10$~ms and atom velocity $v=10$~m/s, this is sufficient to produce a continuous gyroscope with short-term rotation rate sensitivity \hl{$\Delta \Omega \approx 5 \times 10^{-10}$~\si{rad/s}/\si{\sqrt{Hz}}} and accelerometer with short-term acceleration noise\hl{\hbox{ $\Delta a \approx 10^{-8}$~\si{m/s^2}/\si{\sqrt{Hz}}}} at the QPN limit with ideal contrast. Similarly, a Ramsey microwave clock based on the continuous cold-atom beam as in Ref.~\cite{devenoges_improvement_2012}, again assuming ideal fringe contrast and with free evolution time $T=20$~ ms, would have short-term frequency instability at the QPN limit of $\sigma_y \approx 3 \times 10^{-14}$~\si{(\tau / \s)^{-1/2}}. Other implementations such as a CPT beam clock \cite{vanier_atomic_2005,elgin_cold-atom_2019} and optical clock \cite{martin_compact_2018} are expected to be possible using the continuous cold atom beam.

To obtain high fringe contrast $C$, low atom temperatures in three dimensions are desired for a number of reasons. In Raman-based atom interferometers, transverse cooling is beneficial because it reduces the inhomogeneity in Doppler shifts, Rabi frequencies and wavefront \cite{schkolnik_effect_2015} of Raman transitions and also increases interference fringe contrast in the presence of sensor rotation by reducing the inhomogeneous Coriolis acceleration of the atoms \cite{mcguirk_sensitive_2002,stockton_absolute_2011}. Cooling in the longitudinal direction further reduces susceptibility to contrast loss under sensor rotations for the same reason. Longitudinal cooling is also of particular importance in continuous-beam atom interferometers: the value of $T=\ell/v_x$ is determined by the atomic longitudinal velocity $v_x$, such that an inhomogeneous distribution of $v_x$ leads to a distribution of scale factors relating atom interferometer phase to rotation rate and acceleration. This longitudinal broadening of scale factors can be compensated almost perfectly in the case of rotations \cite{gustavson_rotation_2000}, but only approximate compensation is possible for accelerations. As a result, three-dimensional cooling of the atom beam is an effective method of increasing the dynamic range of accelerations and rotation rates that can be measured by a continuously operating atom interferometer. 

The lowest atom temperatures obtained in the present work, $15$~\si{\micro K}, are well below the Doppler cooling limit, but remain higher than the lowest temperatures commonly obtained in pulsed, rather than continuous, polarization gradient cooling of rubidium \cite{gerz_temperature_1993}. The transit time of the atoms in the OM region, equal to approximately 2.5~ms for atoms moving with longitudinal velocity of 11~m/s, should be sufficient to reach temperatures in the few~\si{\micro K} range according to semiclassical cooling models \cite{nienhuis_operator_1991}. The continuous nature of the PGC increases the technical challenge of reaching very low atom temperatures: temperature optimization by varying the temporal profile of key parameters - cooling beam power, cooling beam frequency, and magnetic field - is challenging in practice. These parameters must instead be controlled and varied over the spatial trajectory of the atom beam. We therefore speculate that the atom temperatures obtained are limited primarily by uncontrolled magnetic field gradients and suboptimal spatial profiles of OM intensity and frequency. Further optimization of these parameters is likely to improve the atom beam temperature in three dimensions to values closer to those typical of pulsed PGC experiments. \hl{The addition of a cooling methodology based on pumping into dark states, such as grey molasses \cite{rosi_2018} or degenerate Raman sideband cooling \cite{di_domenico_laser_2004}, would further reduce the final temperature without adding substantially to the cooling-induced fluorescence, but at a cost of additional complexity. Alternatively, two-stage cooling incorporating a high-capture-velocity stage followed by a low-capture-velocity stage with low scattering rate has been demonstrated in alkaline earth atoms \cite{bennetts_2017,chen_2019}}.

As Fig.~\ref{fig:extrapcontrast} shows, for the chosen 3D OM parameters the majority of the residual decoherence due to scattered light in our system comes not from fluorescence emitted by the atom beam, but rather by other spurious reflections and surface scattering. In this initial characterization of the atom beam architecture, no attempt has been made to reduce such reflections using apertures or light baffles at the output of the 3D OM. In contrast, the aperture at the output of the 2D$^+$ MOT eliminates the majority of the non-atomic light scattering. Because the atom beam from the 3D OM is highly collimated, with rms divergence angle $< 5\;\textrm{mrad}$ at the current transverse temperatures, an exit tube with aspect ratio 10:1 or greater could be readily incorporated into the atom beam source, drastically reducing all sources of downstream scattered light apart from atomic fluorescence.

\section{Conclusions}

Here we have presented a continuous, compact, 3D-cooled atom beam source based on a 2-stage laser cooling technique that is expected to improve the prospect for compact, continuous cold-atom interferometry and clocks. The atomic output beam from this device demonstrates temperatures as low as 15~\si{\micro K}, with high flux ($1.6(3)\times 10^9\;\textrm{atoms/s}$). We have studied the sources of quantum decoherence in this system and found that the two-stage continuous cooling technique mitigates decoherence due to atomic fluorescence in the cooling stages. We show that, with little effort to eliminate scattered light from the second cooling stage, quantum coherence can be maintained with a Raman-Ramsey fringe contrast of $80\%$ over a free evolution distance of 20 mm at the beam source output. Further iterations of the vacuum cell and the addition of light baffles following the second cooling stage should further reduce decoherence by an order of magnitude or more, while further improvement of magnetic field and cooling beam profiles are expected to improve atom beam temperatures.

\section{Acknowledgements}

This work was supported in part by the Office of Naval Research. The authors would like to thank Evan Salim of ColdQuanta Inc.\ for valuable discussions and for providing the two-stage vacuum cell used in these experiments. We thank Spencer Olson of Air Force Research Laboratory for providing the experimental control software for this work. This research was performed while the authors JMK and CTF held an NRC Research Associateship award at the U.S. Naval Research Lab.

Author CTF's affiliation with The MITRE Corporation is provided for identification purposes only and is not intended to convey or imply MITRE's concurrence with, or support for, the positions, opinions, or viewpoints expressed by the author.


%

\end{document}